\documentclass[prl,letterpaper,superscriptaddress,twocolumn,showpacs,amsmath,amssymb,floatfix,amstex]{revtex4} 
\usepackage{graphicx,graphics,rotating,bbm}	
\graphicspath{{images/},{figures/}}
\usepackage[utf8]{inputenc}
\usepackage{dcolumn}
\usepackage{color,pstricks}

\def\notext#1{}

\def\etal#1{ {\em et al.}}
\def\tit#1{}

\def\prltext{Supplementary material}

\def\H{{\cal H}} 

\def\ave#1{\langle #1 \rangle}
\def\ii{{\rm i}}
\def\sx{\sigma^{\rm x}}
\def\sy{\sigma^{\rm y}}
\def\sz{\sigma^{\rm z}}
\newcommand{\Van}{{Van\'i\ifmmode \check{c}\else \v{c}\fi{}ek} }
\newcommand{\carlos}[1]{}

\newcommand{\rmi}{\mathrm{i}}
\newcommand{\rme}{\mathrm{e}}
\newcommand{\rmd}{\mathrm{d}}
\newcommand{\eref}[1]{eq.~(\ref{#1})}

\newcommand{\Fref}[1]{Fig.~\ref{#1}}  
	
\newcommand{\mcJ}{\mathcal{J}}	
	
\newcommand{\mcM}{\mathcal{M}}	
\newcommand{\mcU}{\mathcal{U}}	
	
\newcommand{\mcD}{\mathcal{D}}

\newcommand{\env}{\text{env}}	
\newcommand{\cen}{\text{qubit}}	
\newcommand{\GUE}{\text{GUE}}	
	
\newcommand{\tr}{\mathop{\mathrm{tr}}\nolimits}
\newcommand{\diag}{\mathop{\mathrm{diag}}\nolimits}
\newcommand{\<}{\langle}
\renewcommand{\>}{\rangle}
\def\1{{\mathchoice{\rm 1\mskip-4mu l}{\rm 1\mskip-4mu l}{\rm 1\mskip-4.5mu l}{\rm 1\mskip-5mu l} }} 
\newcommand{\jamiol}{Jamio{\l}kowski }

\def\ra{\rangle}
\def\la{\langle}
\usepackage{hyperref}

\definecolor{gris}{rgb}{0.75, 0.725, 0.7}
\definecolor{blood}{rgb}{0.543,0.0273,0.0195} 
\definecolor{pksb}{rgb}{0.137,0.297,0.512}  
\definecolor{verde}{rgb}{0.055,0.543,0.316} 
\definecolor{blood}{rgb}{0.543,0.0273,0.0195}

\begin{document} 
\title{Non-Markovian behavior of small and large complex quantum systems} 
\author{Marko \v Znidari\v c}
\affiliation{Instituto de Ciencias Físicas, Universidad Nacional Autónoma de México, Cuernavaca, México}
\affiliation{Physics Department, Faculty of Mathematics and Physics, University of Ljubljana, Ljubljana, Slovenia}
\author{Carlos Pineda}
\affiliation{Instituto de Física, Universidad Nacional Autónoma de México, México D.F. 01000, México}
\author{Ignacio García-Mata}
\affiliation{\mbox{Instituto de Investigaciones F\'isicas de Mar del Plata (IFIMAR-CONICET), Funes 3350, 7600 Mar del Plata, Argentina}}
\affiliation{Consejo Nacional de Investigaciones Cient\'ificas y Tecnol\'ogicas (CONICET), Argentina}
\affiliation{\mbox{Departamento de F\'isica, Lab. TANDAR --  CNEA , Buenos Aires, Argentina}}
\date{\today}%
\begin{abstract}
The channel induced by a complex system interacting strongly with a qubit
is calculated exactly under the assumption of randomness of its eigenvectors.
The resulting channel is represented as an isotropic time dependent 
oscillation of the Bloch ball, leading to non-Markovian behavior,
even in the limit of infinite environments. 
Two contributions are identified: one due to the density of states and the other 
due to correlations in the spectrum. Prototype examples, one for chaotic and the other for regular dynamics are explored. 
\end{abstract}
%
%
\maketitle
   
{\em Introduction--} Complex quantum systems are of paramount importance in the
description of correlated many-body systems, such as the ones encountered in
condensed matter, as well as few or single body chaotic systems. The exact
description of such complex systems is often not possible because it is either
unfeasible due to many degrees of freedom involved, or impossible because we do
not know all the details of the microscopic model. Frequently we are also
interested only in the dynamics of few degrees of freedom within a larger
system. Unfortunately though, even in this case exact solutions are very rare.
Under certain conditions, which are fulfilled in many important situations, one
can use approximate methods. Such is the case if the central system of interest
is only weakly coupled to the environment with fast decaying correlations. This
leads to the description with a relatively simple Markovian Lindblad master
equation~\cite{Lindblad}, implying a system without memory in which information
flows only out from the central system. While specific models are known in
which the reduced dynamics is not Markovian, general understanding is still
lacking. Such questions resulted in a flurry of recent studies of
non-Markovian behavior~\cite{Jens,Breuer,Plenio,other} and characterization of
reduced dynamics in general~\cite{Wolf,Chruscinski}.
\par 
In the present work we shall derive an exact description of the reduced
dynamics of a single qubit immersed in a complex system, 
undergoing unitary evolution. Our goal is to characterize the one-qubit channel 
induced by this unitary evolution. We shall
assume that the eigenvectors of the Hamiltonian governing such evolution can be
well described by a random unitary matrix. This is a very good approximation if
the system is quantum chaotic~\cite{Haake}, but is also valid under more general
circumstances.
\par
Our main result can be expressed in a very simple geometrical picture.  The
derived one-qubit channel can be imagined as an isotropic shrinking of the
Bloch ball. The radius of this Bloch ball however does not decrease
monotonically with time but instead oscillates, causing non-Markovian behavior.
The oscillations are due to (i) diffraction on the spectral density and, (ii)
due to correlations between eigenenergy levels. Surprisingly, the first
contribution will in general lead to non-Markovian behavior even for an
infinite environment. Comparing the contribution due to eigenenergy
correlations leads us to conclude that in the setting studied, chaotic systems
display stronger non-Markovian behavior than regular ones, as quantified by
measures proposed in~\cite{Breuer,Plenio}.  We also show, via exact
expressions, that the channel is self-averaging for large sizes, meaning that
non-Markovian behavior can be observed in individual system instances.

{\em Setting--} 
We study a system of dimension $N$, divided into a single qubit and the rest,
acting as an environment to which the qubit is strongly coupled. The evolution
of the total system is determined by a Hamiltonian $H$. The only requirement on
$H$ is that the statistical properties of its eigenvectors are described by a
random unitary matrix, which is connected to a maximum entropy
principle~\cite{balian}. This is conjectured to happen for chaotic systems in
the semiclassical limit, and is true, by construction, for the random matrix
ensembles~\cite{RMT} suitable for describing statistical properties of quantum
chaotic systems~\cite{Haake}. In quantum information language we want to characterize the quantum channel
acting on the qubit. Once this is done we can study, for instance, whether the
channel is markovian or not.

Assume that the initial state of the system is a factorizable state, with a
projector in the environment. Other choices of initial states will be discussed
later. The state at later times is thus simply 
\begin{equation}
\rho_\cen^{(t)} = \tr_\env \left[
		U^t \rho_\cen^{(0)} \otimes |\psi_\env\>\<\psi_\env| U^{-t} 
	\right],
\label{eq:rhoevolution}
\end{equation}
where $U^t=\exp{(-\ii H t)}$ (we set $\hbar=1$). This induces a
completely positive map $\rho_\cen^{(t)} =
\Lambda^{(t)}(\rho_\cen^{(0)})$.  The matrix representation of
this linear map in the  basis of Pauli matrices is simply 
\begin{equation}
\Lambda^{(t)}_{j,k} = 
	\frac{1}{2} \tr\left[\sigma^j \Lambda^{(t)}(\sigma^k)\right], 
\label{eq:Lambda}
\end{equation}
where 
$i,j=0,\ldots,3$ with $\sigma^j=\{\sx,\sy,\sz,\mathbbm{1}\}$.

{\em Analytic derivation--} 
We are interested in obtaining explicit expressions for \eref{eq:Lambda}.
Writing $H$ in its eigenbasis as $H= W \diag(E_i) W^\dagger$, where $W$ is the
unitary matrix of eigenvectors of $H$, we are interested in properties of
$\Lambda^{(t)}$ for a unitarily invariant ensemble of Hamiltonians where $W$ is
a random unitary matrix. We shall calculate the average values of all matrix
elements of channel $\Lambda^{(t)}$ as well as its fluctuations. One finds that
given the invariance of $H$ under unitary rotations, the average channel, ie.,
after averaging over the unitarily invariant Haar measure of $W$, denoted by
$\la \cdot \ra_\mcU$, must acquire a diagonal form in the Pauli basis (which
can also be checked by an explicit calculation). Such channel is called 
depolarizing channel in quantum information. The matrix
$\la\Lambda^{(t)}_{j,k}\ra_\mcU$ is therefore diagonal with time-dependent
elements 
\begin{equation}
\alpha(t):=
	\ave{\Lambda^{(t)}_{0,0}}_\mcU =  
	\ave{\Lambda^{(t)}_{1,1}}_\mcU =  \ave{\Lambda^{(t)}_{2,2}}_\mcU.
\end{equation}
Trace preservation means that $\ave{\Lambda^{(t)}_{3,3}}_\mcU =1$ and
$\ave{\Lambda^{(t)}_{3,j=0,1,2}}_\mcU=0$. All the physical information about
the average channel, like the presence of non-Markovian effects, is contained
in $\alpha(t)$, which is the radius of the evolved Bloch sphere of the qubit. The
calculation of $\alpha(t)$ proceeds by separating the dependence of $U^t$ on
the spectra $E_i$ and the eigenbasis $W$, $U^t = W \diag \exp(-\rmi E_i t)
W^\dagger$, to obtain
\begin{multline}
\alpha(t)
=  
\rme^{-\rmi (E_i-E_j)t} 
\big\langle
W_{0\mu, i} W_{00, i}^* W_{1\mu, j}^* W_{10, j}  \\+
W_{0\mu, i} W_{10, i}^* W_{1\mu, j}^* W_{00, j}  
\big\rangle_\mcU
,
\nonumber
\end{multline}
where we have used both Einstein's summation convention and tensorial notation.
Latin indices run over the whole system, whereas Greek ones run over the system
minus the qubit. One can then average over the unitary Haar measure of $W$
using the exact formulas in~\cite{Collins}, obtaining the exact expression
\begin{equation}
\alpha(t)=
	\frac{N^2 |f(t)|^2-1}{N^2-1},
\label{eq:L00}
\end{equation}
with $f(t)=\frac1N \sum_j{\exp{(-\ii E_j t)}}$ being the Fourier transform of
the level density. The details of the calculation are to be found in the
additional material~\cite{Supplement}.

The evaluation of the fluctuations of matrix elements is of interest, as 
it indicates how {\it a single} member of the ensemble will resemble the 
behavior of the ensemble average. Its calcultion involves 8-point correlations of $W$,
and the  Weingarten function for permutations on $4$ elements, which we 
have calculated~\cite{Weingarten}.
Let us define by $\sigma^2_{j,k}
= \ave{[\Lambda^{(t)}_{j,k}]^2}_\mcU - \ave{\Lambda^{(t)}_{j,k}}_\mcU^2$
the standard deviation of matrix element $\Lambda^{(t)}_{j,k}$.  Again, due to the
symmetry there are only three different fluctuations: those of diagonal matrix
elements, those of off-diagonal elements in a $3\times 3$ block
$\Lambda^{(t)}_{j,k}$ and those of $\Lambda^{(t)}_{i,3}$.  The
exact expressions to all orders in $1/N$ is given in the additional
material~\cite{Supplement}, here we only give the leading terms in $1/N$, which
are
\begin{align}
\sigma^2_{i,i}=\sigma^2_{i,3}
	&=\frac{1+(f^*(t)^2 f(2t)+f(t)^2 f^*(2t))-3|f(t)|^4}{2N}, \nonumber \\
\sigma^2_{i,j \ne i}
 	&=\frac{1+|f(t)|^4-(f^*(t)^2f(2t)+f(t)^2 f^*(2t))}{2N},
\label{eq:fluctmain}
\end{align}
with $i,j=0,1,2$. Equations (\ref{eq:L00}) and (\ref{eq:fluctmain}) constitute our main result.

In the above results we have taken the initial state of the environment to be a
projector. Due to the unitary invariance we can choose for
$|\psi_\env\>\<\psi_\env|$ any state. Because $\ave{\Lambda^{(t)}_{j,k}}_\mcU$
is linear in the initial state, any convex sum of projectors, ie., a density
matrix of the environment, will also lead to the same average channel.
Fluctuations though, which are not linear in the initial state, do change. In
particular, the size of the fluctuations will scale as $\sim 1/(N\,r)$, if $r$
is the rank of the initial state of the environment. For instance, if the
initial state of the environment is an identity matrix, corresponding to the
environment at high temperature, the fluctuations scale as $\sim 1/N^2$ instead
of $\sim 1/N$ as for the projector, meaning that self-averaging is stronger. 


{\em A random matrix example--} 
We illustrate the above results by taking $H$ from the Gaussian Unitary
Ensemble (GUE).
This kind of Hamiltonians have been successfully used to describe a wide range
of physical systems including chaotic systems, condensed matter systems and
quantum environments \cite{RMT, Haake, pinedalong}. For $t\to \infty$ the
induced channel $\Lambda^{(t)}$ is closely related to the so-called random
quantum channel, in which $U^t$ is replaced by a random unitary. Random quantum
channels are usefull in quantum information theory~\cite{Ion} and have been
used to prove that the conjecture about superaditivity of channel capacities is
false~\cite{aditivity}. Because the joint probability distribution of
eigenvalues is known for GUE we can perform explicit averaging over the
spectrum, obtaining an expression for the average $f(t)$; note that due to
self-averaging for large $N$ the average behavior is observed also in
individual samples. As $\Lambda^{(t)}$ is quadratic in $f(t)$ it can be
expressed in terms of 1- and 2-point correlations, which are known exactly for
any dimension~\cite{RMT}. Strength of the interaction is fixed by $\< |H_{i,j}|^2 \> = \frac{1}{N}$, resulting in the spectral span of $4$ (determining the shortest time scale) and the Heisenberg time being $2N$ (giving the longest time scale, i.e., the inverse level spacing). The level density is
$R_1(E) = 
\sum_{j=0}^{N-1} \varphi_j^2(E)$,
where $\varphi_j(x)=\frac{{\rm e}^{-N x^2/4}}{\sqrt{2^j j! \sqrt{2\pi/N} }}
\H_j(x\sqrt{N/2})$ and $\H_j$ are Hermite polynomials. The cluster function, giving correlations between different levels, is
for GUE $T_2(E_1,E_2)=\left(\sum_{j=0}^{N-1} 
\varphi_j(E_1) \varphi_j(E_2)\right)^2$. One can show that $N^2\ave{|f(t)|^2}_\GUE=N+\int \rmd E_1 \rmd E_2 \rme^{-\rmi(E_1-E_2)t } [R_1(E_1)R_1(E_2)-T_2(E_1,E_2)]$,
which can be evaluated explicitly for any $N$. 
Let us define $b_2(t)=(1/N)
\int \rmd E_1 \rmd E_2 \rme^{-\rmi(E_1-E_2)t }
T_2(E_1,E_2)$ and
$b_1(t)=(1/N)
\int \rmd E  \rme^{-\rmi E t } R_1(E)$.
Normalization is such that $b_1(0)=1$ and $b_2(\infty)=0$. The final formula is
\begin{equation}
\ave{\alpha(t)}_{\rm GUE}=\frac{
N^2 b_1^2(t)+N (1-b_2(t))-1}{N^2-1}.
\label{eq:GUE}
\end{equation}
Each of the contributions approach a simple expression in the limit $N\to \infty$: $\lim_{N \to \infty} b_1(t)=\frac{J_1(2t)}{t}$, while the leading order of the form factor $1-b_2(t)$ is $t/2N$ for $t<2N$ and $1$ otherwise.
\notext{
\begin{eqnarray}
b_1(t)
&
\xrightarrow{\ \text{large } N \ }& \frac{J_1(2t)}{t} \\
1-b_2(t)
&
\xrightarrow{\ \text{large } N \ }&  
\left\{ \begin{array}{rl}
t/2N & \mbox{ if $t< 2N$} \\
1 & \mbox{ otherwise}
\end{array}\right.,
\label{eq:asymptotic}
\end{eqnarray}
where $J_1(t)$ is the Bessel function.}

\begin{figure} 
\includegraphics[angle=-90,width=0.49\textwidth]{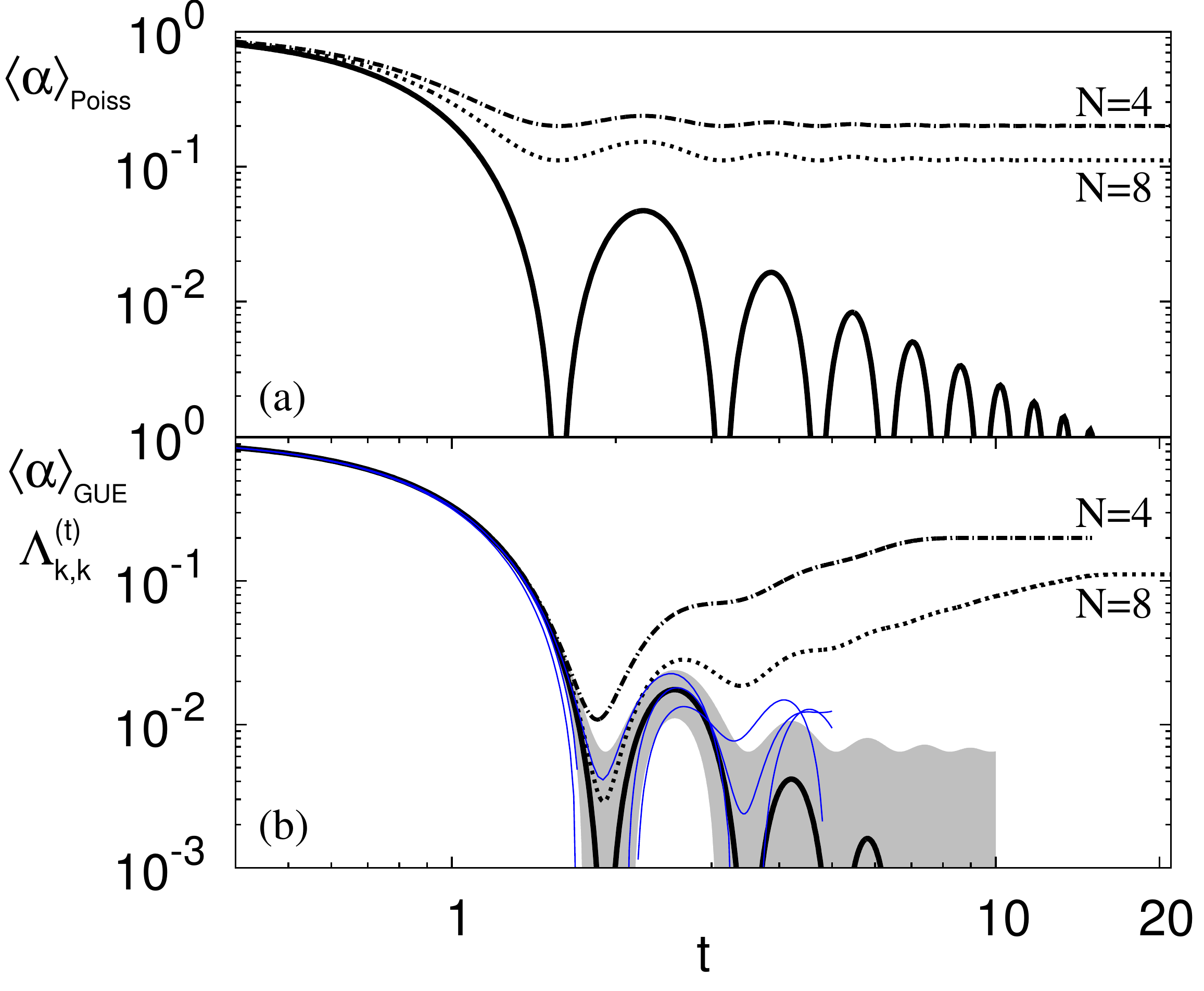} 
\caption{
(a) Theoretical dependence of $\alpha(t)$ for Poisson example, Eq.(\ref{eq:alphapoi}) for $N=4,8$ and $\infty$. (b) Same three sizes for GUE example, Eq.(\ref{eq:GUE}). We also show three diagonal elements of $\Lambda^{(t)}$ for one instance of $N=12000$ (three thin blue curves) and theoretial fluctuations, displayed around $N=\infty$ theory as a gray shadow.}
\label{fig:both}
\end{figure} 

We can see that there are two contributions to $\ave{\alpha(t)}_{\rm GUE}$. The
first one comes from the Fourier transformation of the energy density. The second one, given by the form factor and being due to eigenenergy correlations, is of the order $1/N$ compared to the
first one, and is therefore important for moderate $N$. For large $N$ the second
term can be neglected in Eq.~\ref{eq:GUE}, giving $\lim_{N \to \infty}
\ave{\alpha(t)}_{\rm GUE}=[J_1(2t)/t]^2$. The form of $\ave{\alpha(t)}_{\rm GUE}$ for small and large $N$ in shown in Fig.~\ref{fig:both}b . There we also show fluctuations, which can be for
large $N$ obtained by using $f(t) \approx J_1(2t)/t$ in Eq.~\ref{eq:fluctmain}
. The fluctuations decay with the system size as $\sim 1/N$. Therefore, for
sufficiently large system the fluctuations are smaller than
$\ave{\alpha(t)}_{\rm GUE}$, i.e., the dynamics is self-averaging. Even taking
a single member $H$ of the GUE ensemble one gets the average behavior
$\ave{\alpha(t)}_{\rm GUE}$, as can be seen in Fig.~\ref{fig:both} for
$N=12000$. 
If we would take the initial state of the environment to be
the maximally mixed state,
instead of $|\psi_\env\>\<\psi_\env|$, similar
self-averaging would be achieved already for $N \approx \sqrt{12000} \approx
100$, which is about $7$ qubits.

{\em Poisson example--} 
As a second example we show one still possessing unitary invariance, but having Poissonian eigenenergies with no correlations~\cite{seligmanpoe}, and with a flat level density, being a model for regular systems. The calculation goes exactly as in the previous example. Taking into account that there are no correlations among different levels and the spectral density is flat ($T_2(E_1, E_2)=0$, $R_1(E) = N\Theta\left(\left| E-2\right|\right)/4$, where $\Theta$ is the Heaviside step function), we get (see \Fref{fig:both}a)
\begin{equation}
\langle \alpha(t)\rangle_\text{Poisson}=\frac{N}{N+1}\left[ \frac{\sin(2t)}{2t} \right] ^2+\frac{1}{N+1}.
\label{eq:alphapoi}
\end{equation}

{\em Non-Markovian behavior--} 
Having calculated $\alpha(t)$, one can  immediately draw conclusions about the
non-Markovian behavior of the channels. Consider the map that takes a state from a
time $t$ to $t+\tau$, $ \Lambda^{(t,
t+\tau)}=\Lambda^{(t+\tau)} \left[ \Lambda^{(t)} \right]^{-1}$. This is, in
general, not a physical map, which implies that the trace one operator
associated via the \jamiol isomorphism $\mcJ$  is not a physical state.  In
\cite{Plenio}, the deviation of positivity for such operators is taken as a
measure of non-Markovian behavior $\mcM_1$. We define
\begin{equation}
g(t) \equiv \lim_{\epsilon \to 0^+}\frac{||\mcJ(\Lambda^{(t+\epsilon, t)} ) ||_1 -1}{\epsilon}
=
\left\{
\begin{array}{cl}
\frac{3\dot \alpha(t)}{2\alpha(t)}  & \mbox{ if $\dot \alpha(t) > 0$} \\
0 & \mbox{ otherwise}
\end{array}\right.
\label{eq:plenio}
\end{equation}
which will be positive whenever $\alpha(t)$ increases (the details are
presented in the supplementary material~\cite{Supplement}).  With this figure of merit one can
calculate the values of $ \mcM_1= \int_0^\infty g(t) \rmd t $.
A different criterion is based on the evolution of distinguishability of states
with time~\cite{Breuer} and is defined as $\mcM_2=\max_{\rho_{0,
1}(0)}\int_{\sigma > 0}\rmd t \sigma(\rho_0, \rho_1, t)$, where 
$\sigma(\rho_0(t), \rho_1(t))$ is derivative of the trace distance 
between $\rho_{0,1}(t)$.  The states $ \rho_{0,
1}$ that maximize such quantity for our channel are any two orthogonal pure states, say $\rho_i
=|i\>\<i|$. In such case $\rho_i(t)=\frac{\openone \pm \alpha(t) \sigma_x}{2}$
and $\mcM_2 = 2 \int_{\dot \alpha > 0}\rmd t \dot \alpha(t)$. The last measure
to be examined quantifies 
non-Markovian behavior via the non-monotonicity of
entanglement decay of our qubit with an ancilla qubit~\cite{Plenio} and is as such, as we will see, weaker than $M_{1,2}$:
$\mcM_3=\int_{\dot C>0} \rmd t [\rmd C(\rho(t))/\rmd t]$, where $C$ is a
measure of entanglement (to be taken here as the concurrence~\cite{concurr}), $\rho(0)$ is a
Bell state in the two qubits and the quantum channel acts on a single qubit.
The concurrence for the corresponding state will be in our case
$C(\rho(\alpha))= \max\{0, (3\alpha-1)/2\}$. The final result is $\mcM_3 =
\frac32 \int_{\dot \alpha>0, \alpha>1/3} \dot \alpha \rmd t$.  In
table~\ref{tab:highlander} we report several values of all three measures for
different environments. We can see that both $\mcM_{1, 2}$ indicate non-Markovian behavior exactly at times when
$\alpha(t)$ increases, in other words, when the Bloch ball expands. If we explicitly write $\mcM_1=3/2 \sum_{}{\ln{(\alpha(t_f))}-\ln{(\alpha(t_i))}}$ and $\mcM_1=2 \sum_{}{\alpha(t_f)-\alpha(t_i)}$, where both summations are over all intervals $[t_i,t_f]$ on which $\alpha(t)$ increases, it is also easy to understand why the behavior of $\mcM_{1,2}$ is different with $N$. Because of the divergence of logarithm at $0$, the behavior of $\mcM_1$ is dominated by values of $\alpha(t_i)$ which decrease with $N$, eventually becoming $0$ for $N \to \infty$, causing the increase of $\mcM_1$ with $N$. On the other hand $\mcM_2$ is dominated by terms $\alpha(t_f)$ that decrease with $N$, see Fig.~\ref{fig:both}b . For Poisson example $\mcM_2$ increases due to a trivial $N/(N+1)$ prefactor. Looking back at our results and the two examples of a GUE and Poissonian ensemble, we
can see that for small times non-Markovian behavior is due to diffraction on the spectral density. 
Provided the spectral span $\Xi$ is finite, there will {\it always} be oscillations in
$\alpha(t)$ on the time-scale $1/\Xi$, causing non-Markovian behavior. How fast these oscillations decay with time depends on the singularity at
the spectral edge -- sharper features lead to slower decay of oscillations with
time. In condensed matter systems singularities at spectral edges (van Hove
singularities) are quite common. Surprisingly, non-Markovian behavior is present even
for an infinite environment, where one would perhaps expect that there is no
``back-flow of information'' from the environment to the qubit. For smaller
systems the term with the 2-point correlations also leads to non-Markovian
effects. Indeed, for chaotic systems $1-b_2(t)$ increases with time, leading to
an additional increase of $\alpha(t)$. This contribution occurs on the
time-scale of the inverse level spacing. Interesting to note is, that comparing
the GUE case, mimicking chaotic systems, with the Poissonian for regular
dynamics, shown in Fig.~\ref{fig:both} , one can conclude that non-Markovian
effects are stronger in chaotic systems than in regular ones. This is
yet-another example of a counter intuitive behavior of quantum chaotic system.
Another is their stability, where quantum chaotic systems can be less
sensitive to perturbations than regular ones~\cite{fid}.     
\begin{center}
\begin{table}[ht!]
\begin{tabular}{|l|c|c|c|c|c|c|}\hline
& \multicolumn{3}{c|}{GUE}& \multicolumn{3}{c|}{Poisson}\\ \hline
 & $\mcM_1$ & $\mcM_2$& $\mcM_3$& $\mcM_1$&  $\mcM_2$&  $\mcM_3$\\
\hline \hline
$N=4$         & $4.375$  & $0.378$ & $0$ & $0.555$   & $0.156$   & $0$\\ \hline
$N=8$         & $6.102$  & $0.236$ & $0$ & $1.064$   & $0.173$   & $0$\\ \hline
$N\to \infty$ & $\infty$ & $0.051$ & $0$ & $\infty$  & $0.195$   & $0$\\ \hline
\end{tabular}
\caption{Different values of non-Markovian behavior for several environments.
Notice how the two measures $\mcM_{1, 2}$ have different tendency for the 
GUE case, and how $\mcM_3$ can not be used to detect non-Markovian behavior in our systems. 
}\label{tab:highlander}
\end{table}
\end{center}
{\em Conclusion--} 
We analytically calculate a quantum channel describing the reduced dynamics of
a single qubit within a larger system. Unitary evolution by unitarily invariant
Hamiltonian \carlos{Some people read only conclusions. A unitaryly invariant hamiltonian 
is too vague. Any ideas on how ot reformulate?} 
leads to simple diagonal channel that can be visualized as an
isotropically oscillating Bloch ball. The average value of the diagonal matrix
element has two contributions: (i) one from the Fourier transformation of the
energy density, and (ii) from correlations between eigenenergies. Provided
there is some eigenenergy repulsion, as is the case in quantum chaotic systems,
the second contribution will lead to semiclassically small non-Markovian behavior. This effect is stronger for more chaotic systems. The contribution due to energy
density in general leads to non-Markovian effects even in the limit of an
infinite environment. We also calculate channel fluctuations, showing that the
dynamics is self-averaging for large systems. This means that non-Markovian
effects should be observable already in small individual systems, making it an
exciting experimental challenge.  
{\em Acknowledgments--} Support by the Program
P1-0044, the Grant J1-2208 of the Slovenian Research Agency,  and projects
CONACyT 57334 and  UNAM-PAPIIT IN117310 are acknowledged.


\section{\prltext{}} 
\subsection{Fluctuations of $\Lambda_{j,k}^{(t)}$} 

\begin{figure}
\centering
\includegraphics[angle=-90, width=\columnwidth]{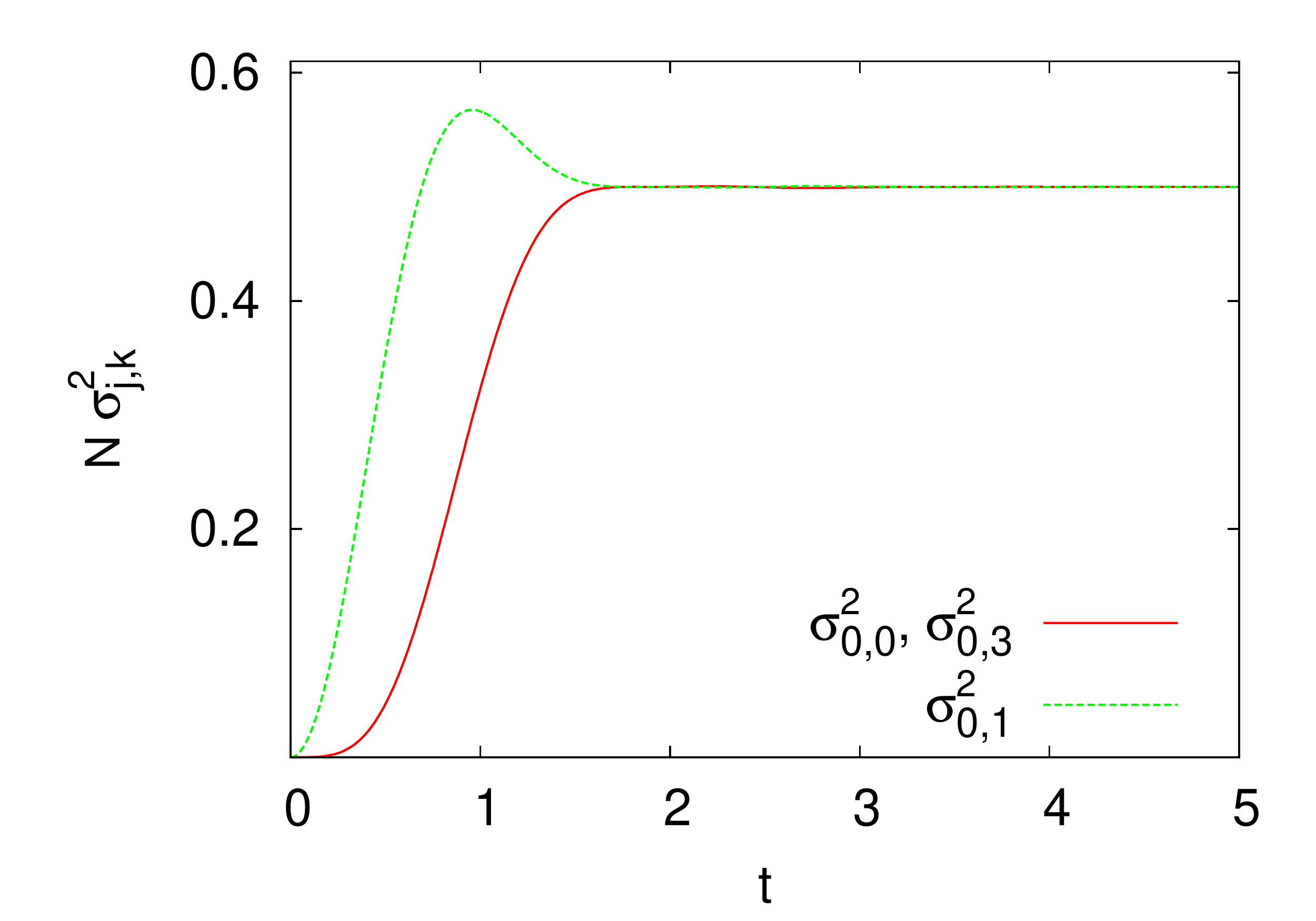}
\caption{ 
Scaled fluctuations of matrix elements of $\Lambda^{(t)}$ for a GUE ensemble and large sizes, where one can use $h_t$ for $f(t)$.
}
\label{fig:sigmaNinfty}
\end{figure}

\begin{figure}
\centering
\includegraphics[angle=-90, width=\columnwidth]{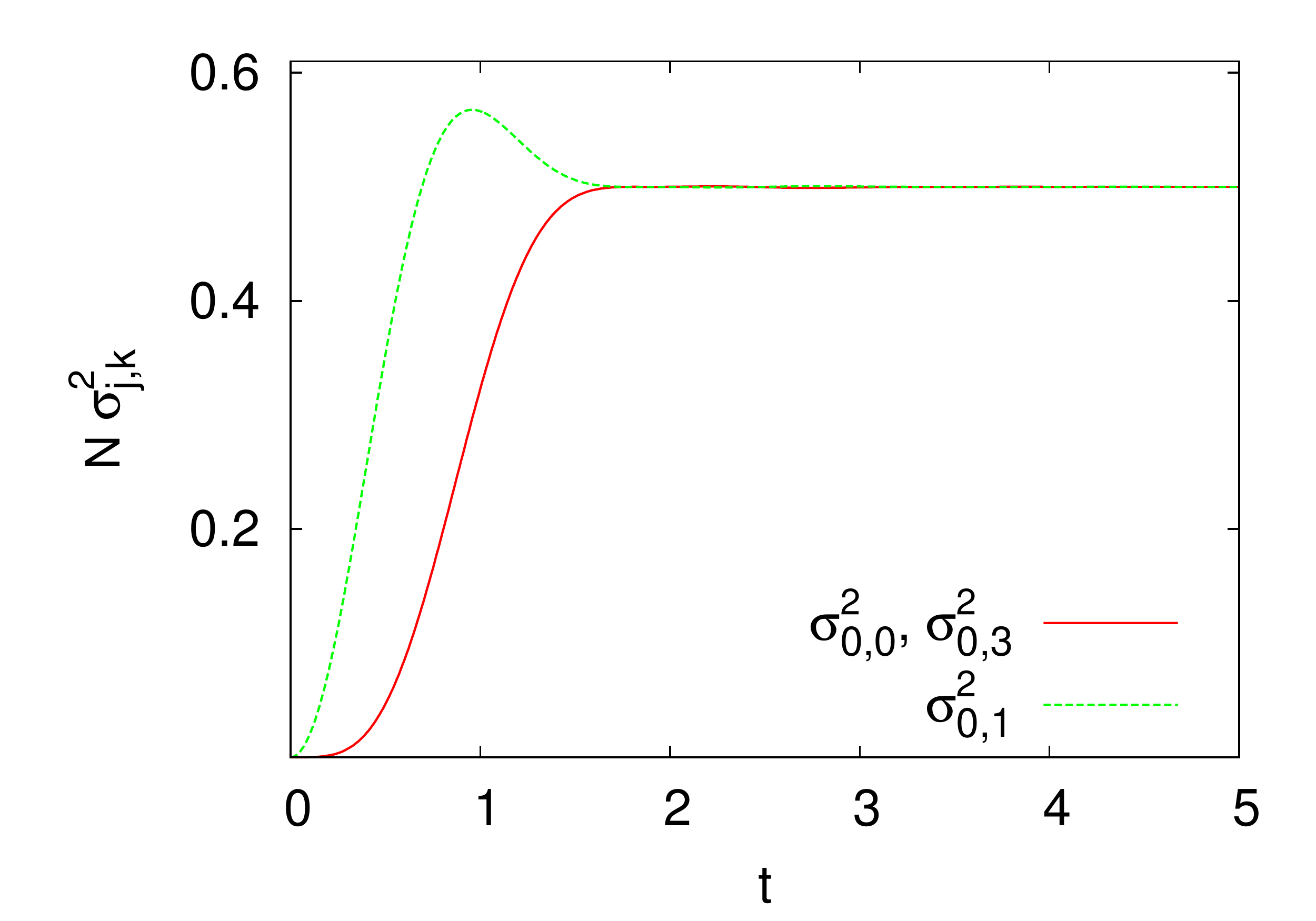}
\caption{ 
Values of off-diagonal elements of $\Lambda^{(t)}_{j,k}$, with $j\neq k=0,1,2$, for one GUE instance of size $N=12000$ (thin blue curves). Fluctuation $\sigma_{0,1}$ is shown as a gray shadow around the average at $0$ (thick red line).
}
\label{fig:off}
\end{figure}

\begin{figure}
\centering
\includegraphics[angle=-90, width=\columnwidth]{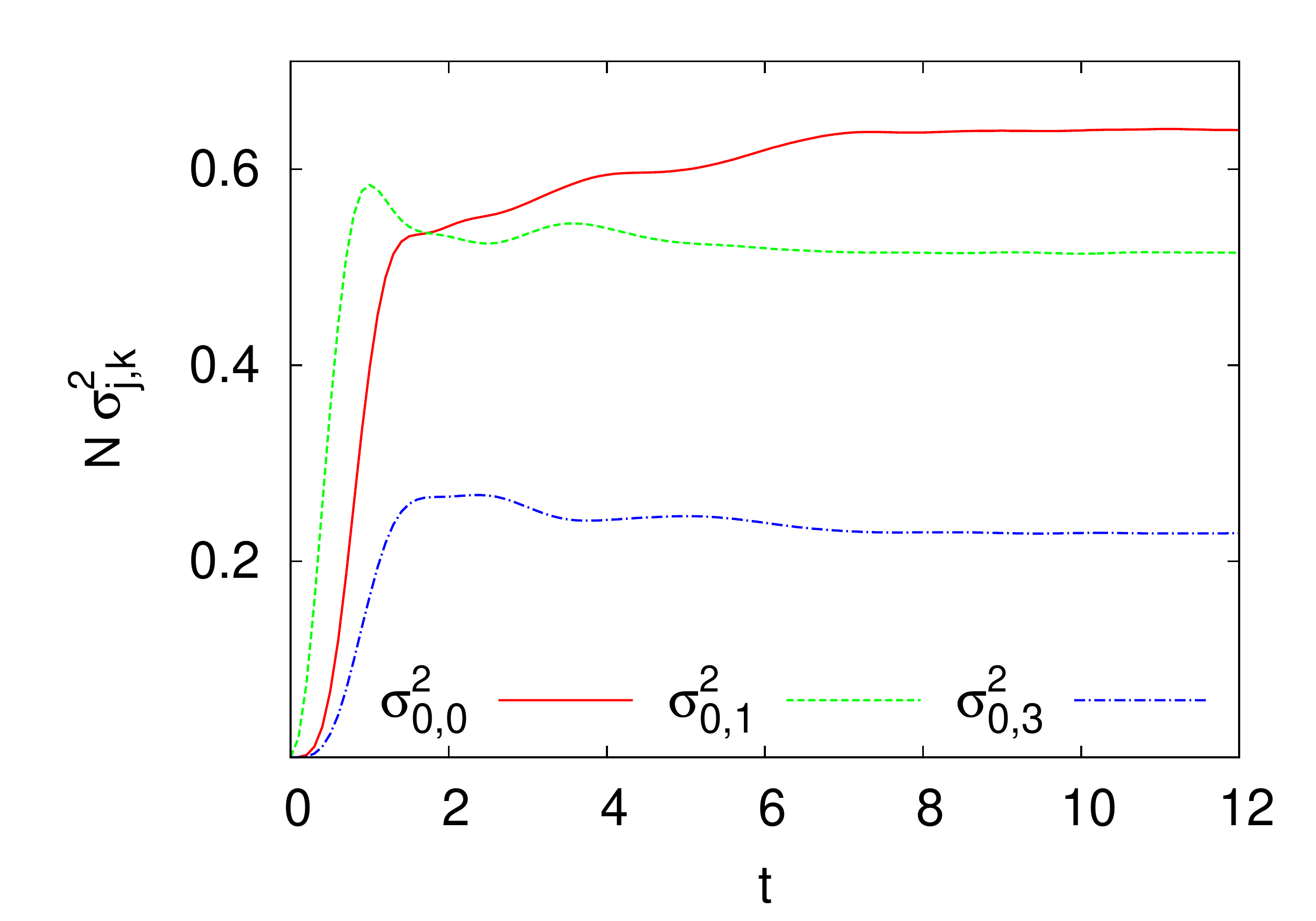}
\caption{
Theoretical formulas for fluctuations given in Eqs.(\ref{eq:fluct00},\ref{eq:fluct03},\ref{eq:fluct01}) for size $N=4$. All is for a GUE ensemble.
}
\label{fig:sigmaN4}
\end{figure}

After straightforward but tedious calculation,
we obtained exact results for all three different fluctuations. They can all be
expressed in terms of the Fourier transformation of the level density denoted
by $f(t)$ and its powers,
\begin{equation}
f(t):=\frac{1}{N}\sum_j \exp{(-\ii E_j t)}.
\label{eq:f}
\end{equation}
Exact expressions (to all orders in $1/N$) for fluctuations of matrix elements of $\Lambda^{(t)}$ are the following (in these expressions no averaging over eigenenergies is performed yet; therefore, any spectrum can be used):
\begin{widetext}
\begin{equation}
\begin{split}
\ave{[\Lambda_{0,0}^{(t)}]^2}_\mcU= \frac{1}{2N(1-\frac{1}{N^2})(1-\frac{9}{N^2})} & \left[   (1-\frac{9}{N^2})+ (2-\frac{3}{N}-\frac{6}{N^2})\left\{ N |f(t)|^4 +\frac{1}{N}|f(2t)|^2-\frac{4}{N} |f(t)|^2 \right\}+ \right. \\
& \left. + (1-\frac{4}{N})\left\{ [f^*(t)]^2 f(2t)+[f(t)]^2 f^*(2t) \right\} \right].
\end{split}
\label{eq:fluct00}
\end{equation}
\end{widetext}
To get the fluctuation, one has to subtract from the previous expression $\ave{\Lambda^{(t)}_{0,0}}^2$ (here the averaging over both, unitary and eigenenergies, has to be performed before squaring). Other two fluctuations are
\begin{widetext}
\begin{equation}
\sigma^2(\Lambda_{0,3}^{(t)})= \frac{(1-\frac{2}{N})}{2N(1-\frac{1}{N^2})(1-\frac{9}{N^2})}  \left[  (1-\frac{9}{N^2})-3|f(t)|^4 -\frac{3}{N^2}|f(2t)|^2+\frac{12}{N^2} |f(t)|^2 + \left\{ [f^*(t)]^2 f(2t)+[f(t)]^2 f^*(2t) \right\} \right].
\label{eq:fluct03}
\end{equation}
\begin{equation}
\begin{split}
\sigma^2(\Lambda_{0,1}^{(t)})= \frac{1}{2N(1-\frac{1}{N^2})(1-\frac{9}{N^2})} & \left[ (1-\frac{9}{N^2})+|f(t)|^4 +\frac{1}{N^2}|f(2t)|^2-\frac{4}{N^2} |f(t)|^2 - \right. \\
 &  \left. -(1-\frac{6}{N^2})\left\{ [f^*(t)]^2 f(2t)+[f(t)]^2 f^*(2t) \right\} \right].
\end{split}
\label{eq:fluct01}
\end{equation}
\end{widetext}
Notice that
one gets correlations up to 4th order, including those that mix $f(t)$ at different
times, making the exact averaging over eigenenergies, for instance in terms of Hermite polynomials for GUE, very cumbersome. In the leading order with respect to $1/N$ one  can, often, forget about
correlations, and interchange averages of powers  with the powers of averages.
Thus, if $h_t:=\ave{f(t)}_{\rm spectrum}$, we obtain for the leading order
\begin{eqnarray}
\sigma^2(\Lambda^{(t)}_{0,0})&=&\frac{1+(h_{t}^{*2} h_{2t}+h_{t}^2 h_{2t}^*)-3|h_t|^4}{2N}, \nonumber \\
\sigma^2(\Lambda^{(t)}_{0,1})&=&\frac{1+|h_t|^4-(h_{t}^{*2} h_{2t}+h_{t}^2 h_{2t}^*)}{2N},\nonumber \\
\sigma^2(\Lambda^{(t)}_{0,3})&=&\sigma^2(\Lambda^{(t)}_{0, 0}).
\label{eq:fluct}
\end{eqnarray}
This approximation is indeed valid in both examples examined in the 
main text for large dimensions. For GUE ensemble $h_t=J_1(2t)/t$, giving very simple expression for correlations. They are shown in Fig.~\ref{fig:sigmaNinfty} . In Fig.~\ref{fig:off} we also show the values of 6 off-diagonal matrix elements $\Lambda^{(t)}_{j,k}$, $j\neq k=0,1,2$, for one GUE instance of dimension $N=12000$ (the same data as shown in the main text). Because the average values of these off-diagonal elements is zero (thick line at $0$ in the figure), they simply fluctuate around $0$ with the amplitude given by theoretical $\sigma_{0,1}$, Eq.(\ref{eq:fluct}), and shown as a gray shadow in the figure.
For smaller size $N=4$, and again GUE ensemble, theoretical expressions for fluctuations (\ref{eq:fluct00},\ref{eq:fluct03},\ref{eq:fluct01}) are shown in Fig.~\ref{fig:sigmaN4} . One can see that the time dependence is quite complicated. 
General feature of fluctuations is that they are very small for short times, and reach their maximal value before the first revival in $\alpha(t)$ (this comes about due to the presence of $f(2t)$ term in fluctuations).

\subsection{Details for the GUE calculation} 
We want to evaluate the quantity 
\begin{equation}
\left\langle |f(t)|^2 \right\rangle_\GUE = \frac{1}{N^2}
\left\langle \sum_{i,j} \rme^{-\rmi(E_i-E_j)t} \right\rangle_\GUE,
\label{eq:def}
\end{equation}
where averaging is over GUE spectrum. Such average can be written as 
\begin{multline}
\left\langle \sum_{i,j} \rme^{-\rmi(E_i-E_j)t} \right\rangle_\GUE = 
N + 
\\
\int \rmd E_1 \rmd E_2 \rme^{-\rmi(E_1-E_2)t }
\left\langle \sum_{i \ne j} \delta(E_1 - E_i)\delta(E_2-E_j) \right\rangle_\GUE.
\nonumber
\end{multline}
The quantity to be averaged is the two point correlation function:
\begin{equation}
R_2(E_1,E_2) = 
\left\langle \sum_{i\ne j} \delta(E_1 - E_i)\delta(E_2-E_j) \right\rangle_\GUE,
\label{eq:artwo}
\end{equation}
which can be expressed in terms of the level density of states
$R_1(E)$ and the two level cluster function $T_2(E_1, E_2)$:
\begin{equation}
R_2(E_1,E_2) = R_1(E_1)R_1(E_2)-T_2(E_1,E_2),
\label{eq:rt}
\end{equation}
for which explicit expressions in terms of Hermite polynomials exists. 
In particular, 
\begin{equation}
R_1(E) = \sum_{j=0}^{N-1} \varphi^2_j(E),
\end{equation}
and
\begin{equation}
T_2(E_1, E_2) = \left( \sum_{j=0}^{N-1} \varphi_j(E_1)\varphi_j(E_2)
\right)^2.
\end{equation}
with
\begin{equation}
\varphi_j(x)=\frac{{\rm e}^{-N x^2/4}}{\sqrt{2^j j! \sqrt{2\pi/N} }}
\H_j(x\sqrt{N/2}),
\label{eq:appHermite}
\end{equation}
and $\H_j$ being the Hermite polynomials. Explicit expressions for moderate $N$s can be obtained 
by straightforward calculation with the aid of symbolic computational program.
The final expression for the 
desired quantity is thus
\begin{equation}
\left\langle |f(t)|^2 \right\rangle_\GUE = \frac{
N+
\int \rmd E_1 \rmd E_2 \rme^{-\rmi(E_1-E_2)t } R_2(E_1,E_2) 
}{N^2},
\end{equation}
where $R_2$, $R_1$ and $T_2$ are given in terms of the Hermite polynomials.

In the large $N$ limit, simple expressions are also available. 	The limit of the 
level density is known as the semicircle law, and yields an ellipse with semi axis
determined by normalization. Its Fourier transform is a Bessel function $J_1$:
\begin{equation}
\int_{-\infty}^{\infty} \rmd E R_1(E) \rme^{-\rmi E t}
\xrightarrow{\ \text{large} N \ }
 N \frac{J_1(2t)}{t}.
\label{}
\end{equation}
For the second term, the integral in the 
large $N$ limit yields the well known two level form factor for the GUE:
\begin{multline}
\int \rmd E_1 \rmd E_2 \rme^{-\rmi(E_1-E_2)t } 
T_2(E_1, E_2)  \\
\xrightarrow{\ \text{large} N \ }
\left\{
\begin{array}{rl}
N - |t|/2 & \mbox{ if $t< 2N$} \\
0 & \mbox{ otherwise}
\end{array}\right.. 
\end{multline}
\subsection{Measures of non-Markovian behavior} 
The depolarizing channel maps a state 
$\rho=\frac{\openone+\vec r\cdot \sigma}{2} \to \mcD_\alpha(\rho)=\frac{\openone+
\alpha(t)\vec r\cdot \sigma}{2}$. This is precisely the map corresponding to 
$\<\Lambda \>_\mcU$. We shall now work in the Choi basis, which 
is, for a single qubit, $\{ |0\>\<0|, |0\>\<1|,|1\>\<0|,|1\>\<1|\}$.
The matrix representation of such a channel, in the 
aforementioned  basis is 
\begin{equation}
\mcD_\alpha=\begin{pmatrix}
\frac{1+\alpha}2 & 0 & 0 & \frac{1-\alpha}2 \\
0&\alpha&0&0\\
0&0&\alpha&0\\
\frac{1-\alpha}2 & 0 & 0 &\frac{1+\alpha}2 
\end{pmatrix}.
\label{eq:choidepol}
\end{equation}
One can also think of the map from a time $t_1$ to $t_2$, which is not
necessarily physical.  The matrix representation of such a map is 
\begin{equation}
\mcD_{t_2,t_1}=\begin{pmatrix}
\frac12+\alpha_r & 0 & 0 & \frac12-\alpha_r \\
0&\alpha_r&0&0\\
0&0&\alpha_r&0\\
\frac12-\alpha_r & 0 & 0 &\frac12+\alpha_r
\end{pmatrix}
\end{equation}
with $\alpha_r=\alpha(t_2)/\alpha(t_1)$.
The associated state, via the \jamiol isomorphism $\mcJ$ is
\begin{equation}
\mcJ \mcD_{t_2,t_1}=
\frac12\begin{pmatrix}
\frac12+\alpha_r & 0 & 0 & \alpha_r \\
0&\frac12-\alpha_r&0&0\\
0&0&\frac12-\alpha_r &0\\
\alpha_r & 0 & 0 &\frac12+\alpha_r
\end{pmatrix}
\end{equation}
with eigenvalues $(1-\alpha_r)/4$ (three times) and $(1+3\alpha_r)/4$. 
With this one directly arrives to the result.



\begin{thebibliography}{}

\bibitem{Lindblad} V.~Gorini\etal{, A.~Kossakowski,
 and E.~C.~G. Sudarshan}, \tit{Completely positive dynamical semigroups of N-level systems} J.~Math.~Phys. {\bf 17}, 821 (1976); G.~Lindblad, \tit{On the generators of quantum dynamical semigroups} Commun. Math. Phys. {\bf 48}, 119 (1976).

\bibitem{Jens} M.~M. Wolf\etal{, J.~Eisert, T.~S.~Cubitt, and J.~I.~Cirac}, \tit{Assessing non-Markovian quantum dynamics} Phys.~Rev.~Lett. {\bf 101}, 150402 (2008). 

\bibitem{Breuer} H.-P.~Breuer\etal{, E.-M.~Laine, and J.~Piilo}, \tit{Measure for the degree of non-Markovian behavior of quantum processes in open systems} Phys.~Rev.~Lett. {\bf 103}, 210401 (2009).

\bibitem{Plenio} \' A.~Rivas\etal{, S.~F.~Huelga, and M.~B.~Plenio}, \tit{Entanglement and non-Markovianity of quantum evolutions} Phys.~Rev.~Lett. {\bf 105}, 050403 (2010).

\bibitem{other} S.~Daffer\etal{, K.~W\' odkiewicz, J.~D.~Cresser, and J.~K.~McIver}, \tit{Depolarizing channel as a completely positive map with memory} Phys.~Rev.~A {\bf 70}, 010304(R) (2004); S.~Maniscalco and F.~Petruccione, \tit{Non-Markovian dynamics of a qubit} Phys.~Rev.~A {\bf 73}, 012111 (2006); E.-M. Laine\etal{, J.~Piilo, and H.-P.~Breuer}, \tit{Measure for the non-Markovianity of quantum processes} Phys.~Rev.~A {\bf 81}, 062115 (2010); L.~Mazzola\etal{, E.-M.~Laine, H.-P.~Breuer, S.~Maniscalco, and J.~Piilo}, \tit{Phenomenological memory-kernel master equations and time-dependent markovian processes} Phys.~Rev.~A {\bf 81}, 062120 (2010); B.~Vacchini and H.-P.~Breuer, \tit{Exact master equations for the non-Markovian decay of a qubit} Phys.~Rev.~A {\bf 81}, 042103 (2010); T.~J.~G.~Apollaro\etal{, C.~Di Franco, F.~Plastina, and M.~Paternostro}, \tit{Memory-keeping effects and forgetfulness in the dynamics of a qubit coupled to a spin chain} Phys.~Rev.~A {\bf 83}, 03
 2103 (2011); D.~Chru\' sci\' nski\etal{, A.~Kossakowski, and \' A.~Rivas}, \tit{Measures of non-Markovianity: Divisibility versus backflow of information} Phys.~Rev.~A {\bf 83}, 052128 (2011).

\bibitem{Wolf} M.~M.~Wolf and J.~I.~Cirac, \tit{Dividing quantum channels} Commun.~Math.~Phys. {\bf 279}, 147 (2008). 

\bibitem{Chruscinski} D.~Chru\' sci\' nski and A.~Kossakowski, \tit{Non-Markovian quantum dynamics: Local versus nonlocal} Phys.~Rev.~Lett. {\bf 104}, 070406 (2010).

\bibitem{Haake} F.~Haake, {\em Quantum signatures of chaos}, 3rd ed. (Springer, 2010).

\bibitem{RMT} M.~L.~Mehta, {\em Random matrices}, 2nd ed. (Academic Press, New York, 1990); T.~Guhr\etal{, A.~M\" uller-Groeling, and H.~A.~Weinden\" uller}, \tit{Random-matrix theories in quantum physics: common concepts} Phys.~Rep. {\bf 299}, 189 (1998).

\bibitem{Collins} B.~Collins, \tit{Moments and cumulants of polynomial random variables on unitary groups, the Itzykson-Zuber integral, and free probability} 
Int.~Math.~Res.~Not. {\bf 17}, 953 (2003).

\bibitem{Weingarten} The Weingarten function is defined on the symmetric group $S_q$ of $q$ elements. The value depends only on the length of cycles in permutation (its cycle shape). For $q=2$ and $3$ it is given in~\cite{Collins}, for $q=4$ the values are ${\rm Wg}([4])=-\frac{5}{ab}, {\rm Wg}([3,1])=\frac{2N^2-3}{Nab}, {\rm Wg}([2^2])=\frac{N^2+6}{Nab}, {\rm Wg}([2,1^2])=-\frac{N^2-4}{ab}, {\rm Wg}([1^4])=\frac{N^4-8N^2+6}{Nab}$, where $a=(N^2-4)(N^2-9)$ and $b=N(N^2-1)$. 

\bibitem{Supplement} \prltext{}.

\bibitem{pinedalong} C.~Pineda\etal{, T.~Gorin, and T.~H.~Seligman}, \tit{Decoherence of two-qubit systems: a random matrix description} New J.~Phys. {\bf 9}, 106 (2007).

\bibitem{seligmanpoe} F.-M.~Dittes\etal{, I.~Rotter, and T.~H.~Seligman}, \tit{Chaotic behaviour of scattering induced by strong external coupling} Phys.~Lett.~A {\bf 158}, 14 (1991); M.~Moshe\etal{, H.~Neuberger, and B.~Shapiro}, \tit{Generalized ensemble of random matrices} Phys.~Rev.~Lett. {\bf 73}, 1497 (1994).


\bibitem{balian}R.~Balian, \tit{Random matrices and information theory} Il Nuovo Cimento B {\bf 57}, 183 (1968).
 
\bibitem{Ion} B.~Collins and I.~Nechita, \tit{Gaussianization and eigenvalue statistics for random quantum channels (III)} {\tt e-print arXiv:0910.1768}.

\bibitem{aditivity} P.~Hayden and A.~Winter, \tit{Counterexamples to the maximal p-norm multiplicativity conjecture for all $p>1$} Commun.~Math.~Phys. {\bf 284}, 263, (2008); M.~B.~Hastings, \tit{A counterexample to additivity of minimum output entropy} Nature Physics {\bf 5}, 255 (2009).

\bibitem{concurr} S.~Hill and W.~K.~Wooters, \tit{Entanglement of a pair of quantum bits} Phys.~Rev.~Lett. {\bf 78}, 5022 (1997); W.~K.~Wooters, \tit{Entanglement of formation of an arbitrary state of two qubits} Phys.~Rev.~Lett. {\bf 80}, 2245 (1998).

\bibitem{fid} T.~Prosen and M.~\v Znidari\v c, \tit{Can quantum chaos enhance the stability of quantum computation?} J.~Phys.~A {\bf 34}, L681 (2001); T.~Prosen, \tit{General relation between quantum ergodicity and fidelity of quantum dynamics} Phys.~Rev.~E {\bf 65}, 036208 (2002); T.~Prosen and M.~\v Znidari\v c, \tit{Stability of quantum motion and correlation decay} J.~Phys.~A {\bf 35}, 1455 (2002).

\end{thebibliography}
\end{document}